\documentclass[hyper,a4paper]{JHEP3} 
\usepackage{graphicx,graphics}
\usepackage{cite,citesort}
\usepackage{rotating,rotate}
\usepackage{epsfig}

\usepackage{exscale}
\usepackage{amsmath,amssymb}
\usepackage{cite}
\usepackage{multirow}
\usepackage{latexsym}
\usepackage[english]{babel}
\newcommand{\bc}{\begin{center}}
\newcommand{\ec}{\end{center}}
\newcommand{\bfi}{\begin{figure}}
\newcommand{\efi}{\end{figure}}
\newcommand{\bm}{\begin{minipage}}
\renewcommand{\em}{\end{minipage}}
\newcommand{\be}{\begin{equation}}
\newcommand{\en}{\end{equation}}
\newcommand{\bea}{\begin{eqnarray}}
\newcommand{\eea}{\end{eqnarray}}
\newcommand{\bi}{\begin{itemize}}
\newcommand{\ei}{\end{itemize}}
\newcommand{\no}{\nonumber}

\newcommand{\ov}[1]{\overline{#1}}
\newcommand{\bbr}{{\it B{\footnotesize A}B{\footnotesize AR}}}

\def\b         {\ensuremath{\mathcal{B}}}
\def\bb        {\ensuremath{\mathcal{B}\overline{\mathcal{B}}}}
\def\eepp        {\ensuremath{e^+ e^- \!\rightarrow p\overline{p}}}
\def\ppee        {\ensuremath{p\overline{p} \rightarrow e^+ e^- \!}}
\def\eebb        {\ensuremath{e^+ e^- \!\rightarrow \mathcal{B}\mathcal{\overline{B}}}}
\def\eell        {\ensuremath{e^+ e^- \!\rightarrow \Lambda\overline{\Lambda}}}
\def\eess        {\ensuremath{e^+ e^- \!\rightarrow \Sigma^0\overline{\Sigma^0}}}
\def\ees+s+        {\ensuremath{e^+ e^- \!\rightarrow \Sigma^+\overline{\Sigma^+}}}
\def\eels        {\ensuremath{e^+ e^- \!\rightarrow \Lambda\overline{\Sigma^0}}}
\def\ss        {\ensuremath{\Sigma^0\overline{\Sigma^0}}}
\def\ls        {\ensuremath{\Lambda\overline{\Sigma^0}}}
\def\eenn        {\ensuremath{e^+ e^-\!\rightarrow n\overline{n}}}
\def\nn        {\ensuremath{n\overline{n}}}
\def\pp        {\ensuremath{p\overline{p}}}
\def\ll        {\ensuremath{\Lambda\overline{\Lambda}}}
\def\ee        {\ensuremath{e^+e^-}}
\def\nb        {\ensuremath{{\rm nb}}}

\def\gev       {\ensuremath{{\rm GeV}}}

\title{%
Unexpected features of\\
\eepp\ and \eell\ cross sections\\
near threshold 
}
\author{Rinaldo Baldini$^{\rm a,b}$, Simone Pacetti$^{\rm a,b}$, 
Adriano Zallo$^{\rm b}$, and Antonino Zichichi$^{\rm a,d,e}$ \\
 $^{\rm a}$Museo Storico della Fisica e Centro Studi e Ricerche ``E. Fermi'', Rome, Italy \\
 $^{\rm b}$INFN, Laboratori Nazionali di Frascati, Frascati, Italy  \\ 
 $^{\rm d}$INFN and Department of Physics, University of Bologna, Bologna, Italy\\
 $^{\rm e}$CERN, Geneva, Switzerland\vspace{2mm}\\
\hspace{-6mm}\begin{tabular}{ll}
E-mail: &\email{baldini@centrofermi.it}\\
 &\email{simone.pacetti@lnf.infn.it}\\ 
&\email{adriano.zallo@lnf.infn.it}\\
\end{tabular}% 
}

\preprint{arXiv:0711.1725}	% OR: \preprint{Aaaa/Mm/Yy\\Aaa-aa/Nnnnnn}

\abstract{%
Unexpected features of the \bbr\ data on \eebb\ cross sections ($\mathcal{B}$ stands for baryon)
are discussed. These data have been collected, with unprecedented accuracy, by means of the 
initial state radiation technique, which is  particularly suitable in giving  good acceptance 
and energy resolution at threshold. A striking feature observed in the \bbr\ data is the non-vanishing
cross section at threshold for all these processes. This is the expectation  due to the Coulomb 
enhancement factor acting on a charged fermion pair. In the case of \eepp\ it is found that Coulomb 
final state interactions largely dominate the cross section and the form factor is
$|G^p(4 M^2_p)| \sim 1$, which could be a general feature for baryons. In the case of neutral 
baryons an interpretation of the non-vanishing cross section  at threshold is suggested, based
on quark electromagnetic interaction and taking into account the asymmetry between attractive 
and repulsive Coulomb factors. Besides strange baryon cross sections are compared to U-spin 
invariance predictions.
}
\keywords{QCD, Parton Model, Baryon form factors}
\begin{document}

\section{$\sigma(\eebb)$ at threshold}
\indent The significance of baryon time-like form factors (FF) has been pointed out and
looked for in $\ppee$ long time ago~\cite{zichichipp}. However only recently an exhaustive set of data
has been achieved by \bbr, showing unexpected features even if in part predicted on the basis 
of fundamental principles.
Space-like FF behaviors are also driven by basic principles as it was anticipated~\cite{zichichi,rinaldo}, 
but only after thirty years experimentally recognized~\cite{jlab-mitbates}.
Therefore baryon FF's are still a lively topical subject.  

Unexpected features are pointed out in the following, concerning recent 
cross section measurements of  
\be
\eepp \no
\en
and
\be
\eell,\; \ss,\; \ls \no
\en
in the corresponding threshold energy regions. \bbr\ has measured these cross 
sections~\cite{pp,ll} (Fig.~\ref{fig:cross-sections}), 
with unprecedented accuracy, up to an invariant mass of the \bb\ system:
$W_{\bb}~\sim~4$ GeV, by means of the initial state radiation technique (ISR), 
in particular detecting the photon radiated by the incoming beams. 

There are several advantages in measuring processes at threshold in this way:
\begin{itemize}
\item even exactly at the production energy the efficiency is quite high and, in case of 
charged particles collinearly produced, the detector magnetic field provides 
their separation;
\item a very good invariant mass resolution is achieved, $\Delta W_{p\overline{p}} \sim 1$ 
MeV, comparable to what is achieved in a symmetric storage ring; 
\item   a full angular acceptance is also obtained, even at $0^{\rm o}$ and 
$180^{\rm o}$, due to the detection of the radiated photon. 
\end{itemize}
In Born approximation the differential cross section for the process \eebb\ is
\bea
\displaystyle\frac{d\sigma(\ee\!\!\to\bb)}{d\Omega}%(W_{\bb})
\!=\!
\displaystyle\frac{\alpha^2\beta C}{4W_{\bb}^2}\!\left[(1\!+\!\cos^2\theta)
|G_M^\b(W_{\bb}^2)|^2\!+\!\frac{4M_{\b}^2}{W_{\bb}^2}\sin^2\theta|G_E^\b(W_{\bb}^2)|^2\right ],
\label{eq:cross}
\eea
where $\beta$ is the velocity of the outgoing baryon, $C$ is a Coulomb enhancement 
factor, that will be discussed in more detail in the following, $\theta$ is the 
scattering angle in the center of mass (c.m.) frame and, $G_M^\b$ and $G_E^\b$ are the 
magnetic and electric Sachs FF's.
At threshold it is assumed that, according to the analyticity of the Dirac and Pauli FF's
as well as the S-wave dominance, there is one FF only: 
$G_E^\b(4 M^2_\b) = G_M^\b(4 M^2_\b) \equiv G^\b(4 M^2_\b)$.
%
% Cross sections
%
\bfi[ht]
\bc
\epsfig{file=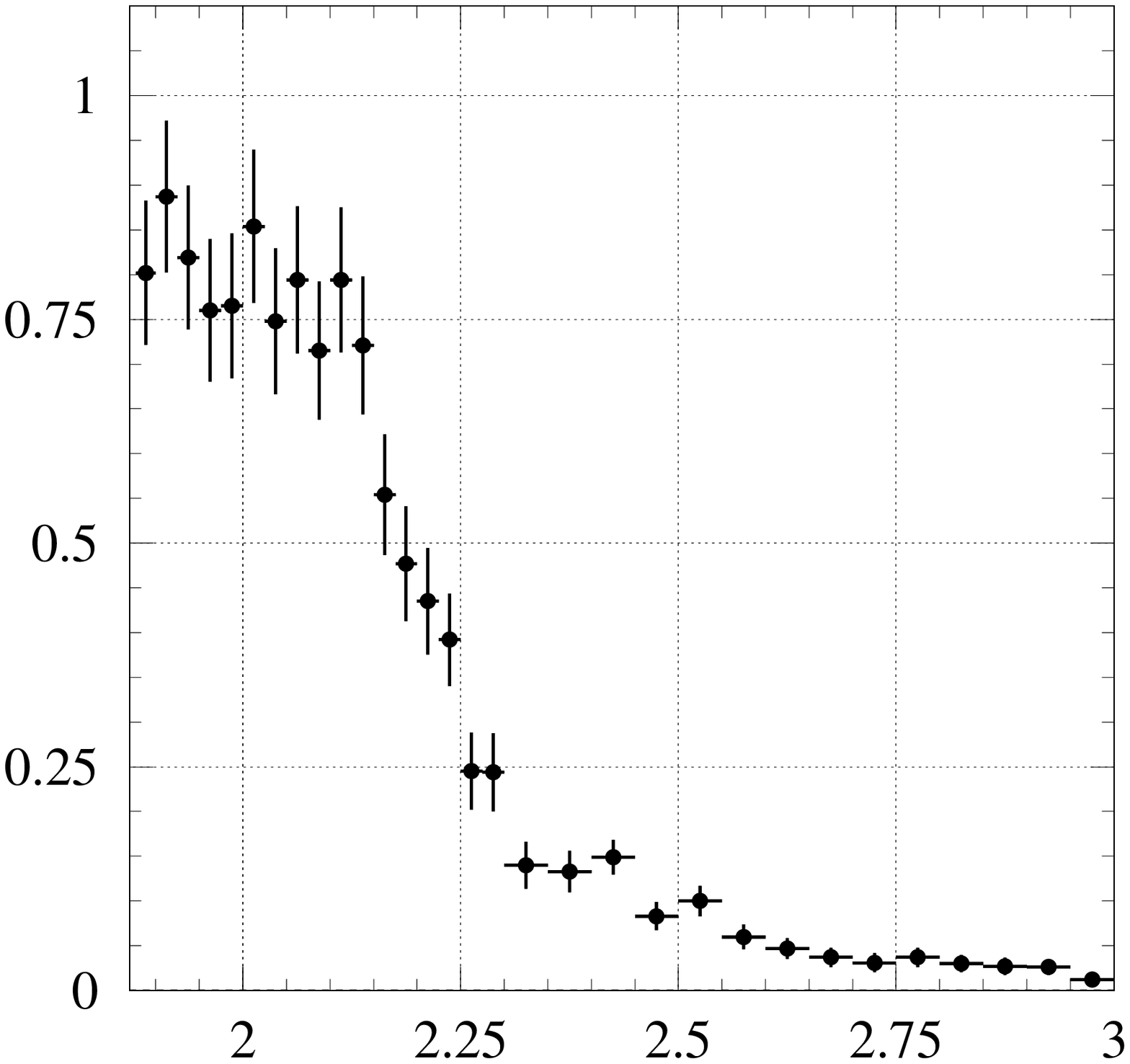,width=65mm}
\put(-57,0){$W_{\pp}$ (GeV)}
\put(-195,92){\rotatebox{90}{$\sigma(\eepp)$ (nb)}}
\put(-25,167){\bf a}
\hspace{10mm}
\epsfig{file=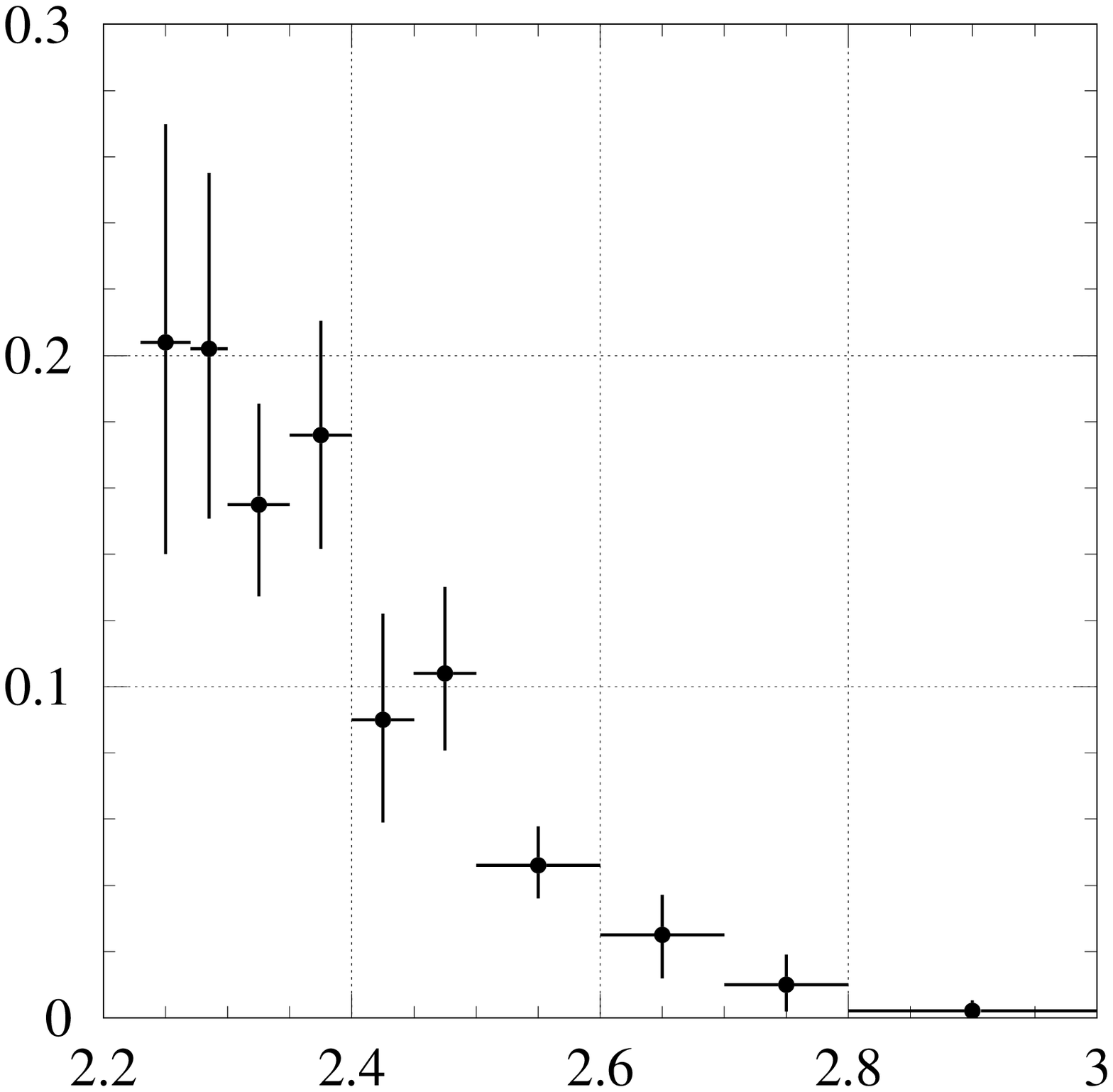,width=65mm}
\put(-61,0){$W_{\ll}$ (GeV)}
\put(-195,88){\rotatebox{90}{$\sigma(\eell)$ (nb)}}
\put(-25,167){\bf b}\vspace{2mm}\\
%%%%%%%
\epsfig{file=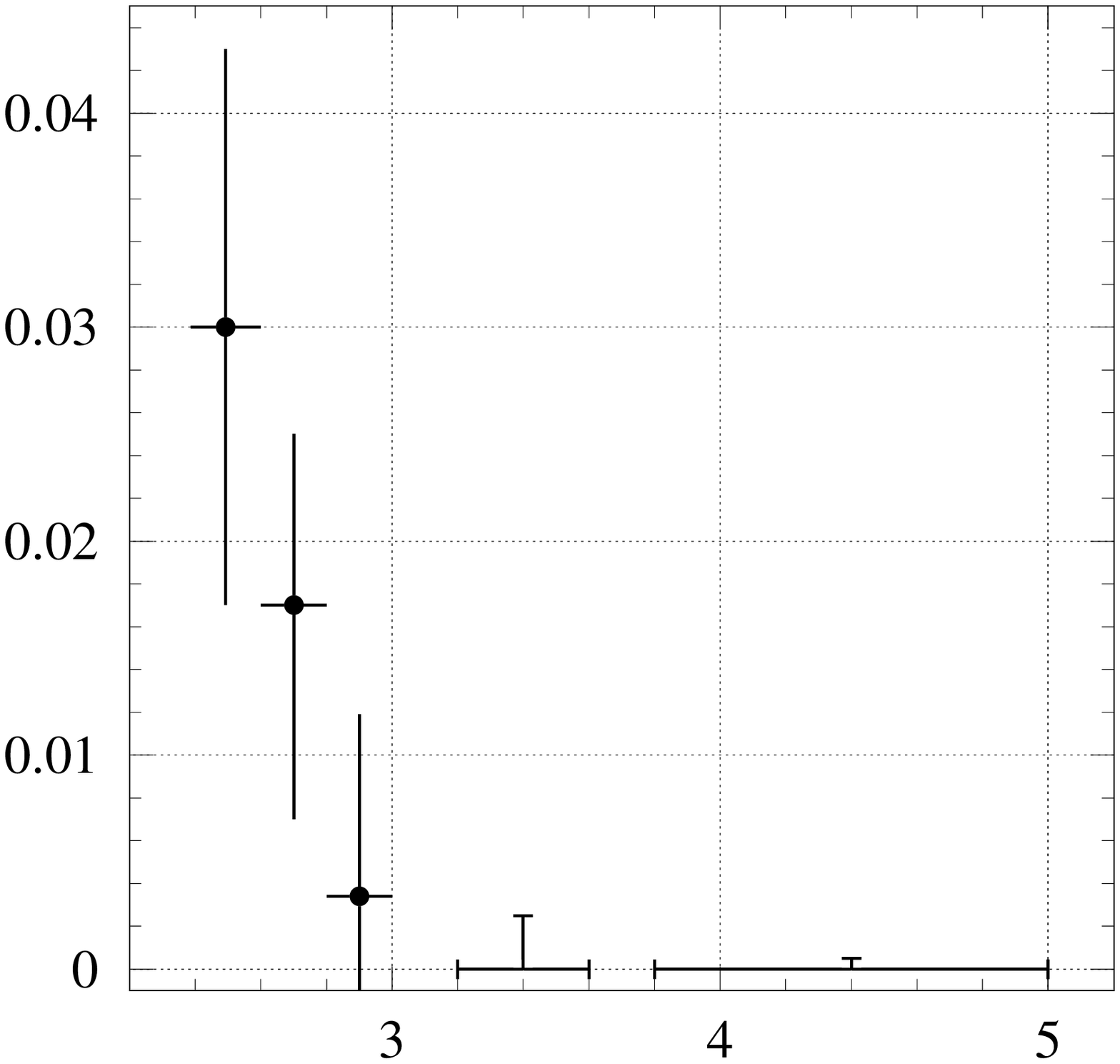,width=65mm}
\put(-69,0){$W_{\ss}$ (GeV)}
\put(-195,77){\rotatebox{90}{$\sigma(\eess)$ (nb)}}
\put(-25,167){\bf c}
\hspace{10mm}
\epsfig{file=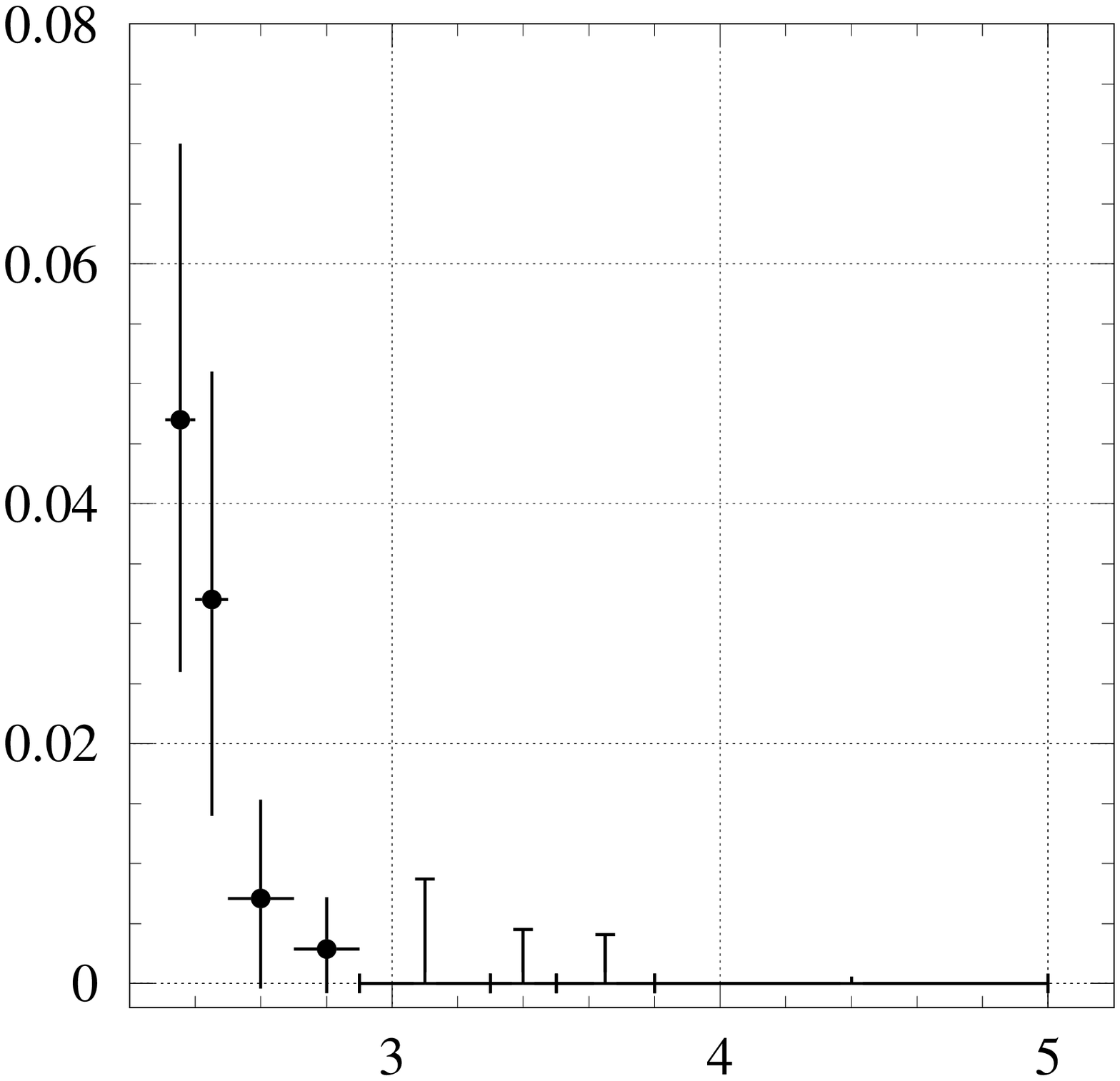,width=65mm}
\put(-65.5,0){$W_{\ls}$ (GeV)}
\put(-195,82){\rotatebox{90}{$\sigma(\eels)$ (nb)}}
\put(-25,167){\bf d}
%%%%%%%
\ec
\vspace{-0mm}
\caption{\label{fig:cross-sections}%
\eepp\ (a), \eell\ (b), \eess\ (c), and \eels\ (d) total cross sections measured 
by the \bbr\ experiment~\cite{pp,ll}.}
\efi

The following peculiar features have been observed, in the case of \eepp~\cite{pp}:
\begin{itemize}
\item as it is shown in Fig.~\ref{fig:cross-sections}a, the total cross section $\sigma(\eepp)$ is suddenly 
different from zero at threshold, being $0.85 \pm  0.05$ \nb\ (by the way it is the only endothermic process 
that has shown this peculiarity);
\item data on $\sigma(\eepp)$ show a flat behavior, within the experimental errors, in an 
interval of about 200 MeV above the threshold and then drop abruptly;
\item the angular distribution, averaged in a 100 MeV interval above the threshold, has a behavior 
like $\sin^2\theta$, i.e. dominated by the electric FF, and then a behavior like 
$(1+\cos^2\theta)$, i.e. dominated by the magnetic FF [see Eq.~(\ref{eq:cross}) and Fig.~\ref{fig:ratio}].
\end{itemize} 
%
% Ratio GE/GM
%
%%%%%%%%%%%%%%
%
\bfi[h]\vspace{-0mm}
\bm{70mm}
\begin{flushright}
\epsfig{file=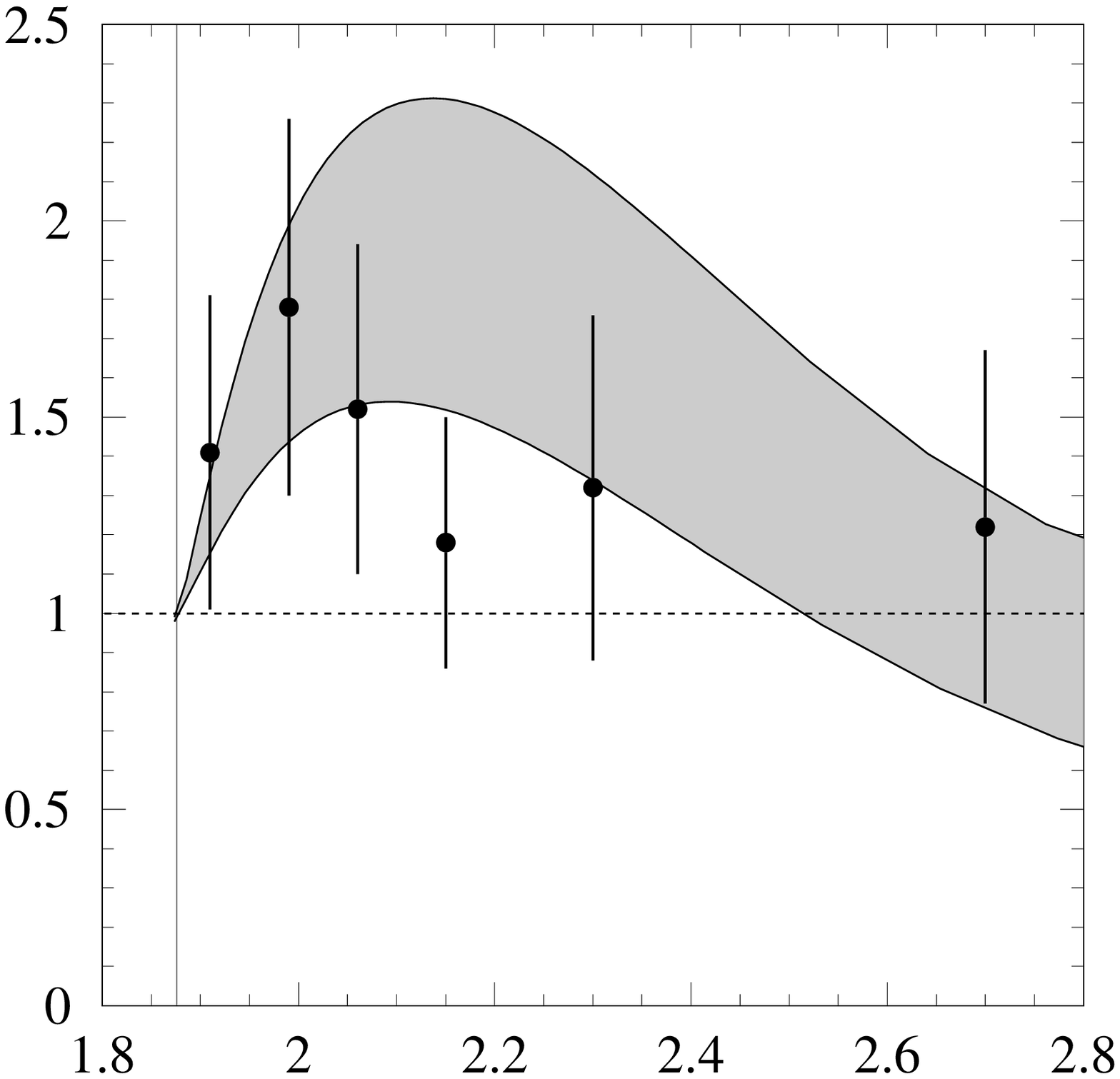,width=65mm}
\put(-53,0){$W_{\pp}(\gev)$}
\put(-195,139){\rotatebox{90}{$|G_E^p/G_M^p|$}}
\caption{\label{fig:ratio}%
\bbr\ data on the ratio $|G_E^p/G_M^p|$ extracted by studying the angular distribution
of the \eepp\ differential cross section [Eq.(\ref{eq:cross})]. The strip is a calculation~\cite{noi} based
on a dispersion relation relating these data and the space-like ratio, as recently achieved
at JLAB and MIT-Bates~\cite{jlab-mitbates}.}
\end{flushright}
\em
%
% Coulomb correction
%
\hspace{5mm}
\raisebox{11.7mm}{
\bm{69mm}
\begin{flushright}
\epsfig{file=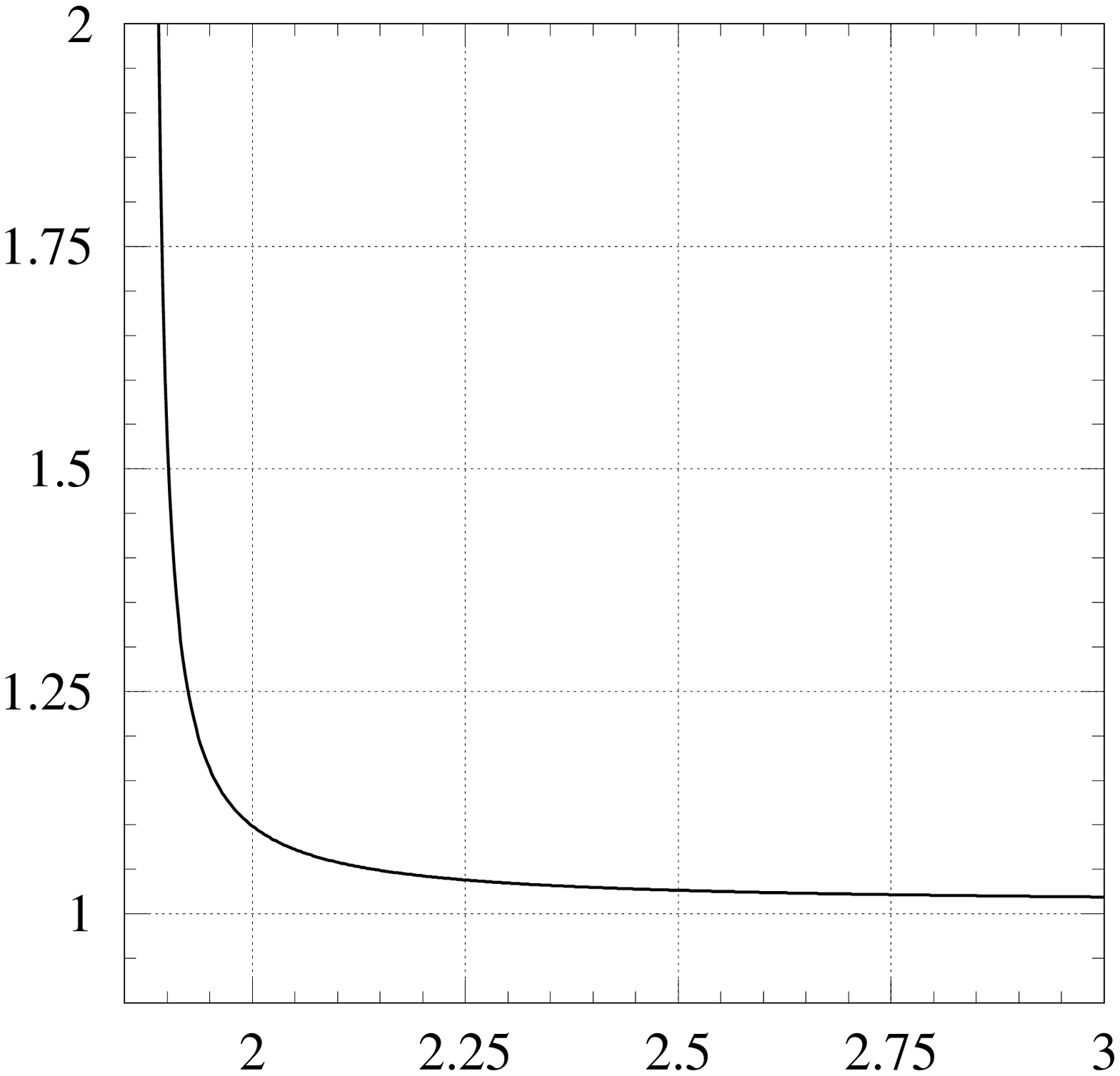,width=65mm}
\put(-53,0){$W_{\pp}(\gev)$}
\put(-195,145){\rotatebox{90}{$C(W_{\pp})$}}
\caption{\label{fig:coulomb-corr}%
Coulomb enhancement factor as a function of the \pp\ c.m. energy from Eq.~(\ref{eq:corr}).}
\end{flushright}
\em}
\efi

Similar features have been observed by \bbr\ in the cases of \eell, \ss, \ls~\cite{ll} 
(Fig.~\ref{fig:cross-sections}b, c, d), even if within much larger experimental errors, in particular 
the cross section  $\sigma(\eell)$ is different from zero at threshold, being $0.20~\pm~0.05$~nb.

Of course, extremely sharp rises from zero cannot be excluded and the relationship between data 
and predictions, reported in the following, could be accidental.

Long time ago it has been pointed out that final state Coulomb corrections to the Born cross section 
have to be taken into account in the case of pointlike charged fermion pair production~\cite{coulomb}.
This Coulomb correction has been usually introduced as an enhancement factor, $C$ in Eq.~(\ref{eq:cross}). 
It corresponds to the squared value of the Coulomb scattering wave function at the origin, assumed as a good 
approximation in the case of a long range interaction added to a short range one, the so called 
Sommerfeld-Schwinger-Sakharov rescattering formula~\cite{coulomb,somm}. This factor has a very weak 
dependence on the fermion pair total spin, hence it is the same for $G_E$ and $G_M$ and can be 
factorized. The Coulomb enhancement factor is
\bea
C(W_{\bb}) = \left\{\begin{array}{ll}
1& \mbox{for neutral $\mathcal{B}$}\\
&\\
\displaystyle\frac{\pi\alpha/\beta}{1-e^{-\pi\alpha/\beta}}& \mbox{for charged $\mathcal{B}$}\\
\end{array}\right., \hspace{10mm}
\beta=\sqrt{\displaystyle1-\frac{4M^2_\mathcal{B}}{W_{\bb}^2}}.
\label{eq:corr}
\eea
In Ref.~\cite{yogi} a similar formula is obtained, but $1/\beta \rightarrow 1/\beta -1$; however
that does not affect the following considerations.
Very near threshold the Coulomb factor is $C(W_{\bb}^2\to 4M_\mathcal{B}^2)\sim\pi \alpha /\beta$,  
so that the phase space factor $\beta$ is cancelled and the cross 
section is expected to be finite and not vanishing even exactly at threshold. 
However, as it is shown in Fig.~\ref{fig:coulomb-corr}, as soon as the fermion relative velocity 
is no more vanishing, actually few MeV above the threshold, it is $C \sim 1$ and Coulomb effects can 
be neglected.

Besides it has been emphasized~\cite{brodsky-cou} that a similar, but quite bigger in amount and 
energy interval, threshold enhancement factor due to strong interactions is forecast in the case 
of heavy quark pair production by $\ee$ annihilation. Low-$Q^2$ gluon exchange should introduce 
in the cross section a factor similar to the Coulomb correction of Eq.~(\ref{eq:corr}), with 
$\frac{4}{3} \alpha_S (Q^2)$ instead of $\alpha$.

In the case of \eepp\ the expected Coulomb-corrected cross section at threshold is
\bea
\sigma(\eepp)(4M_p^2) = \frac{\pi^2\alpha^3}{2M_p^2} \cdot |G^p(4M_{p}^2)|^2  
= 0.85 \cdot |G^p(4 M^2_p)|^2 \;\nb, \no
\eea
in striking similarity with the measured one. 
Therefore Coulomb interaction dominates the energy region near threshold and it is found
\be
|G^p(4 M^2_p)| \sim 1.\no
\en
In the following this feature is suggested
to be a general one for baryons. It looks as if the FF at threshold, interpreted as
$\mathcal{B}$ and $\overline{\mathcal{B}}$ wave function static overlap, 
coincides with the baryon wave function normalization, taking into account S-wave 
is peculiar of fermion pairs at threshold. 
In the case of meson pairs total angular momentum conservation requires a P-wave, 
that vanishes at the origin, hence this Coulomb enhancement factor too, and the 
cross section has a $\beta^3$ behaviour near threshold. Tiny Coulomb effects in the 
case of meson pairs have been extensively pursued~\cite{Voloshin}. 
%
% esempi
%
\bfi[ht]
\bc
\epsfig{file=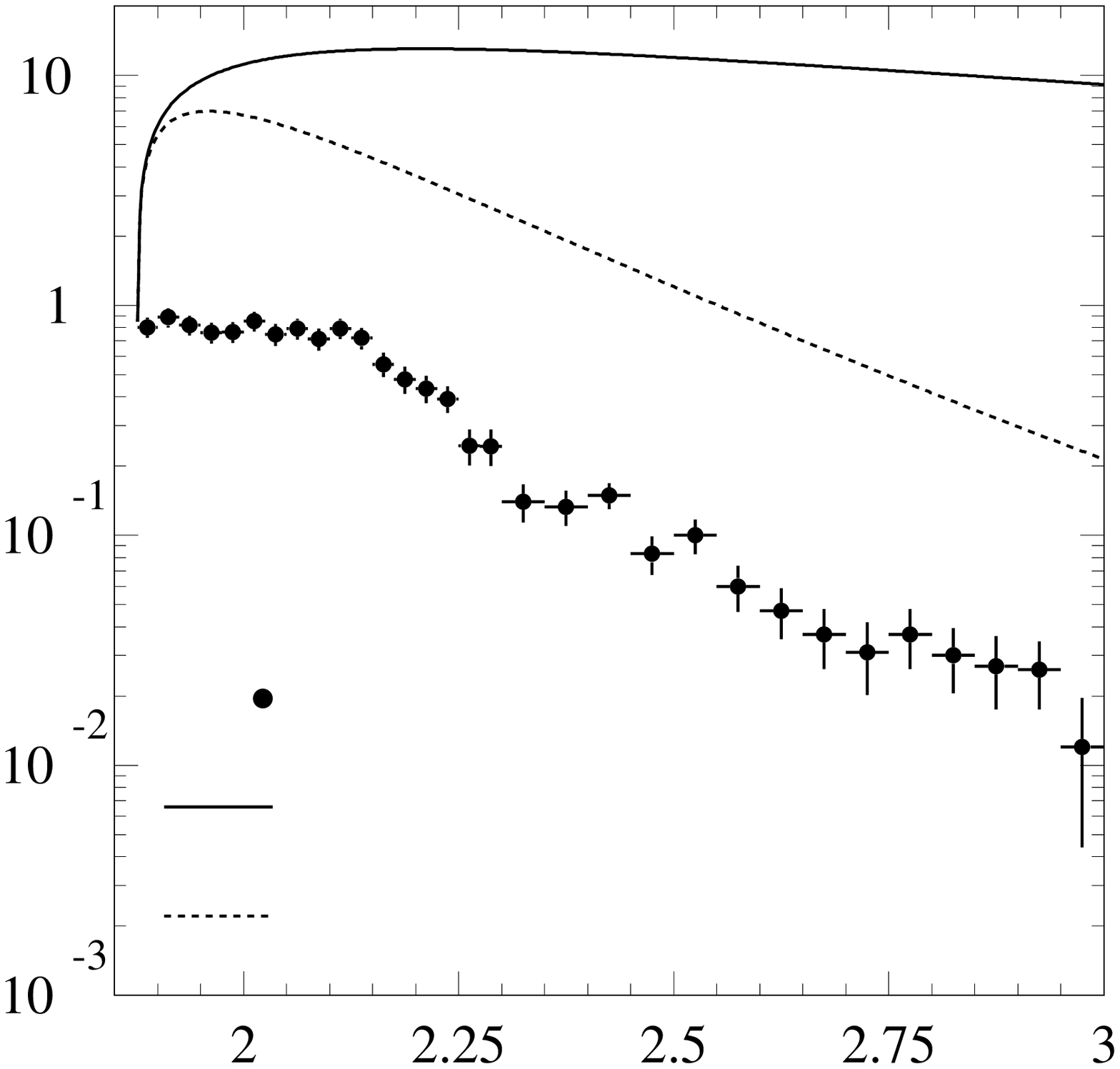,width=90mm}
\put(-53,0){$W_{\pp}(\gev)$}
\put(-263,165){\rotatebox{90}{$\sigma(\eepp)(\nb)$}}
\put(-185,94){\bbr\ data}
\put(-185,73){pointlike proton}
\put(-185,48){$|G^p_{M,E}|\propto 1/W_{\pp}^4$}
\ec
\vspace{-5mm}
\caption{\label{fig:plot1}%
\bbr\ cross section \eepp\ in comparison with expected behaviors in case of
pointlike protons (solid line) and assuming asymptotic FF's (dashed line).
}\efi\\
%
% 

% Cross sections
%
\bfi[ht]
\bc
\epsfig{file=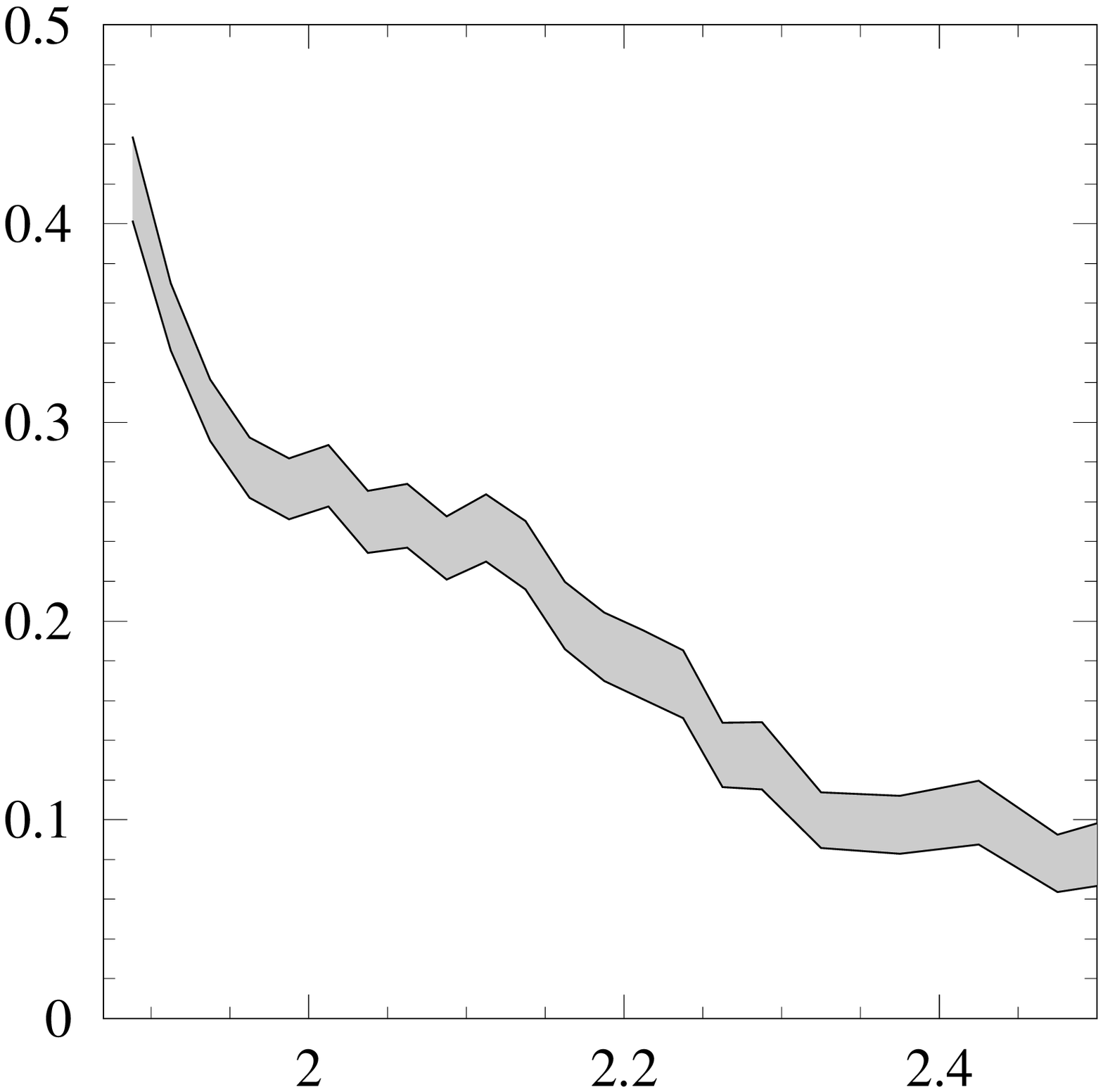,width=65mm}
\put(-57,0){$W_{\pp}$ (GeV)}
\put(-195,136){\rotatebox{90}{$|B_S^p(W_{\pp}^2)|$}}
\put(-20,165){\bf a}
\hspace{10mm}
\epsfig{file=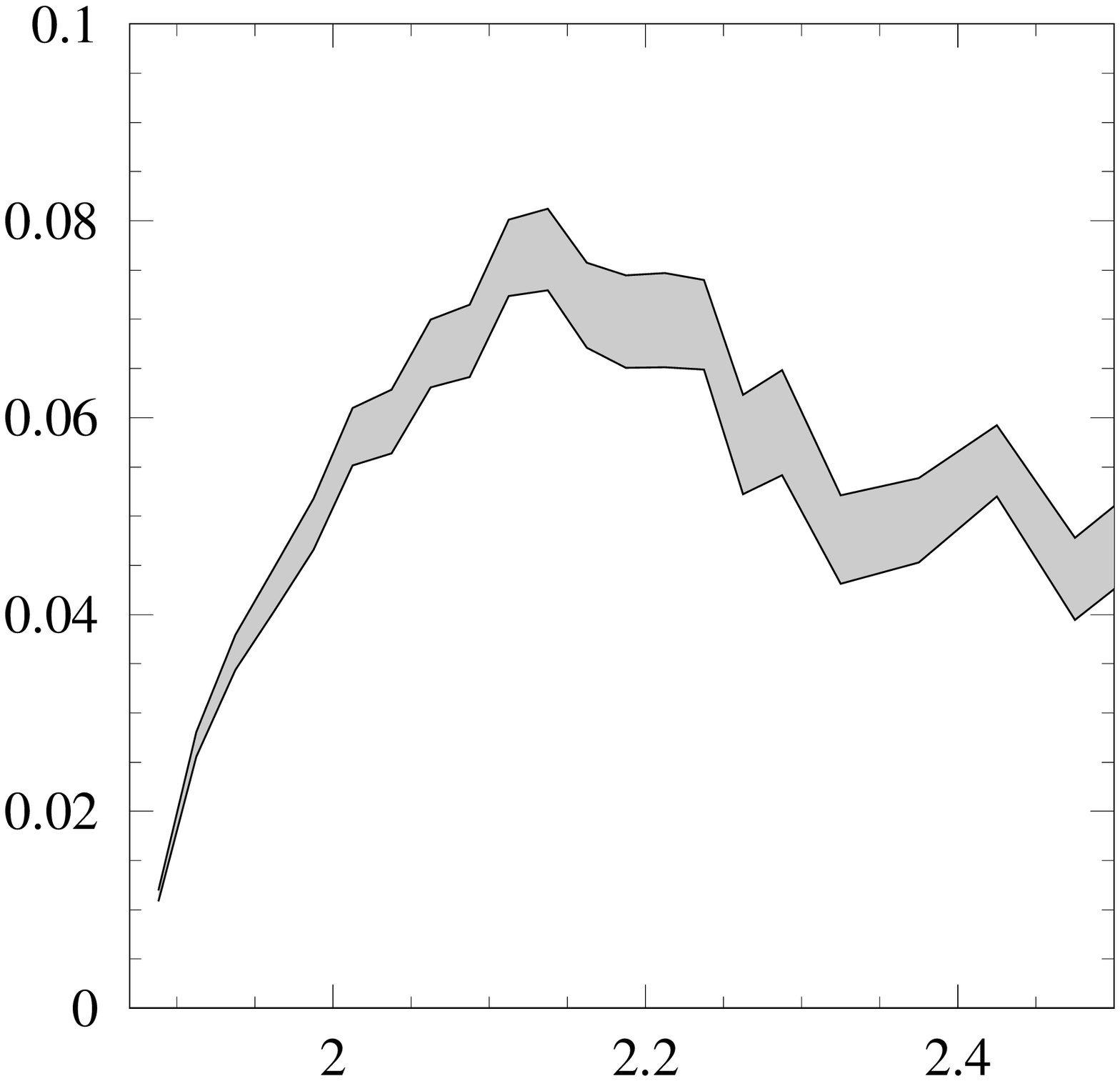,width=65mm}
\put(-57,0){$W_{\pp}$ (GeV)}
\put(-196,134.5){\rotatebox{90}{$|B_D^p(W_{\pp}^2)|$}}
\put(-20,165){\bf b}
\ec
\vspace{-5mm}
\caption{\label{fig:waves}%
S-wave (a) and D-wave (b) FF's as obtained in Ref.~\cite{noi} from a dispersive analysis based
on the \bbr\ data on the total \eepp\ cross section and the time-like ratio $|G_E^p/G_M^p|$.
}
\efi
Why $\sigma(\eepp)$ 
is so flat above the threshold has to be explained as well as the following sharp drop.
As a reference, in Fig.~\ref{fig:plot1} the cross sections, in the case of a pointlike proton 
(solid curve) and in the case of $|G^p_{M,E}| \varpropto 1/W_{\pp}^4$, i.e. $\sigma(\eepp)
\propto 1/W_{\pp}^{10}$ (dashed curve), are shown in comparison with the \bbr\ data. 
A non-trivially structured electric and magnetic FF's [Eq.~(\ref{eq:cross})]
have to be included to get this cross section.
In particular the different behavior at threshold and the dominance of the electric FF are consistent with
a sudden and important D-wave contribution.
In fact, angular momentum and parity conservation allow, in addition to 
the S-wave, also the D-wave contribution. In Ref.~\cite{noi}, by means of a dispersion relation, applied
to the space-like ratio $G_E^p/G_M^p$ and to the \bbr\ time-like $|G_E^p/G_M^p|$
(Fig.~\ref{fig:ratio}), the relative phase and therefore the S- and D-wave complex FF's, 
$B_S^p$ and $B_D^p$, have been extracted. In terms of $G_E^p$ and $G_M^p$ they are:
\bea
B_S^p=(G_M^p W_{\pp}/M_p+G_E^p)/3
\hspace{20mm}
B_D^p=(G_M^p W_{\pp}/2M_p-G_E^p)/3.\no
\eea
S-wave and D-wave  opposite trends, as shown in Fig.~\ref{fig:waves}, produce the observed plateau. 
\section{An interpretation of $\sigma(\eell)$ at the quark level}

In the case of \eell, being $\Lambda$ a neutral baryon, final state Coulomb effects should 
not be taken into account and a finite cross section at threshold is not expected. 
Nevertheless the $\eell$ cross section data (Fig.~\ref{fig:cross-sections}b) show a 
threshold behavior quite similar to that of $\sigma(\eepp)$ (Fig.~\ref{fig:cross-sections}a),
also the ratio $|G_E^\Lambda/G_M^\Lambda|$ (not shown) has a trend similar to
$|G_E^p/G_M^p|$ (Fig.~\ref{fig:ratio}).

Assuming that this Coulomb dominance is not a mere coincidence, one might investigate what is expected at 
the quark level. Valence quarks only are considered in the following. 
The baryon pair relative velocity is equal to the quark pair average relative velocity. 
The quark velocity spread inside the baryon should come mostly from the relative velocity 
among the different quark pairs. Hence for each pair there is a Coulomb attractive 
amplitude times the quark electric charge and each amplitude 
has a phase taking into account the displacement of the quark inside the baryon. 
In addition to the quark pair Coulomb interaction there are contributions from quarks
belonging to different pairs. There are several suppression factors for them: 
relative phase, velocity spread and moreover most of them, coming from quarks having 
charges of the same sign, are repulsive ones. There is no symmetry between  repulsive and attractive 
Coulomb interactions and this asymmetry might explain why there is a non-vanishing cross section at threshold 
even for neutral baryon pairs. In fact in the case of repulsive Coulomb interaction the Sommerfeld 
formula is (charges $Q_{q}$ and $Q_{\ov{q}'}$ have the same sign):
\bea
C(W_{\pp}) =\frac{-\pi\alpha |Q_{q}Q_{\ov{q}'}|/\beta}{1-\exp(+\pi\alpha|Q_qQ_{\ov{q}'}|/\beta)}
\mathop{\longrightarrow}_{W_{\pp}^2\to4M_p^2}0 \no
\eea
i.e. $C=0$ at threshold.
Therefore at the quark level, considering only Coulomb enhancement factors due
to quark pairs, it is expected:
\bea
\sigma(\eepp)(4M_p^2) = \frac{\pi^2\alpha^3}{2M_p^2}(2Q_u^2 + Q_d^2) \cdot |G^p(4M_{p}^2)|^2 
= 0.85 \cdot |G^p(4 M^2)|^2 \,\nb, \no
\eea
in the proton case, and
\bea
\sigma(\eell)(4M_\Lambda^2)= 
\frac{\pi^2\alpha^3}{2M_\Lambda^2}(Q_u^2 + Q_d^2 + Q_s^2) \cdot |G^\Lambda(4M_{\Lambda}^2)|^2 
= 0.4 \cdot  | G^\Lambda(4 M_{\Lambda}^2) |\; \nb, \no
\eea 
in the $\Lambda$ baryon case.

The expectation for \eepp, at quark level as well as at hadron level, is the same, 
namely the total cross section is 0.85 \nb\ (assuming $|G^p(4 M^2_p)|^2 \sim 1$) to be compared to the 
experimental value: $\sigma(\eepp) = 0.85 \pm 0.05$ nb at threshold. 
In the case of $\eell$ the 
expectation range is $(0- 0.4)~\nb$ (still assuming $| G^\Lambda(4 M_\Lambda^2) | \sim 1$) to be 
compared to the experimental value at threshold: $\sigma(\eell)= 0.20 \pm 0.05$ nb.

\section{Other baryon form factor measurements}
\label{sec:other}
The cross sections $\sigma(\eess)$ and $\sigma(\eels)$ have been measured by the 
\bbr\ Collaboration for the first time~\cite{ll}, although with large errors. At threshold,
assuming a smooth extrapolation from the first data point, it is 
$\sigma(\eess)=0.03\pm 0.01$ nb and $\sigma(\eels)=0.047\pm 0.023$ nb. 
The expectation, according to U-spin symmetry and some additional hypotheses on the 
interaction Hamiltonian~\cite{coleman},
is that $\Lambda$ and $\Sigma^0$ have opposite (equal in modulus)
magnetic moments as well as FF's at threshold, apart from mass corrections. Hence, on the basis 
of the $\eell$ cross section it should be 
$\sigma(\eess) \sim \sigma(\eell)\cdot (M_\Lambda /M_{\Sigma^0})^2 \sim 0.18$ nb, 
by far greater than the experimental one.

Although at least
the small mass difference among neutral strange baryons implies small corrections to 
U-spin conservation, full U-spin invariance should hold at enough high $Q^2$. 
A milder version of the U-spin invariance~\cite{park}, obtained under the assumption of negligible 
electromagnetic transitions between U-spin triplet and singlet, like the photon, 
is explored in the following. 
%that is, just the assumption electromagnetic transitions between U-spin triplet 
%and singlet, like the photon, are strongly suppressed.  
Therefore, neglecting $\Lambda$ and $\Sigma^0$ mass difference and extrapolating
the magnetic moment relations to the FF's at threshold,
it should be:
\bea
G_{\Sigma^0}=G_\Lambda-\frac{2}{\sqrt{3}}G_{\ls} \,,
\label{eq:uspin-ff}
\eea
that is, assuming real FF's at threshold or no relative phase
\bea
\sigma_{\ss}=\left[\frac{M_\Lambda}{M_{\Sigma^0}}\sqrt{\sigma_{\ll}}
-\frac{2}{\sqrt{3}}\frac{\ov{M_{\ls}}}{M_{\Sigma^0}}\sqrt{\sigma_{\ls}}\right]^2.
\label{eq:uspin-cs}
\eea
In terms of adimensional quantities, the previous relation can be also written as:
\bea
M_{\Sigma^0}\sqrt{\sigma_{\ss}}-M_\Lambda\sqrt{\sigma_{\ll}}+
\frac{2}{\sqrt{3}}\ov{M_{\ls}}\sqrt{\sigma_{\ls}}=0. \no
\eea
Entering the \bbr\ results
we get the following prediction for the $\sigma_{\ss}$ cross section at
threshold
\bea
\sigma_{\ss}=\left[\frac{M_\Lambda}{M_{\Sigma^0}}\sqrt{\sigma_{\ll}}
-\frac{2}{\sqrt{3}}\frac{\ov{M_{\ls}}}{M_{\Sigma^0}}\sqrt{\sigma_{\ls}}\right]^2=
0.03\pm0.03\,\nb.
\eea
This value, which is quite lower than the $\sigma_{\ll}$ cross section, is 
consistent with the measured one. Using Eq.~(\ref{eq:uspin-cs}) with the \bbr\ data for the
cross sections at threshold
%%%%%%%
\bea
M_{\Sigma^0}\sqrt{\sigma_{\ss}}-M_\Lambda\sqrt{\sigma_{\ll}}+
\frac{2}{\sqrt{3}}\ov{M_{\ls}}\sqrt{\sigma_{\ls}}=(-0.1\pm2.0)\times 10^{-4} \no
\eea
still in agreement with the minimal U-spin invariance prediction, within the 
experimental error.

The asymmetry between $\Lambda$ and $\Sigma^0$ FF's with respect
to the proton case can be settled assuming that a suitable combination is the one properly normalized.
%Assuming it is the U-spin singlet,
%the ansatz is put forward, still neglecting $\Lambda$ and $\Sigma^0$ mass difference:
%\bea
%\frac{1}{4}\left(3G_{\Sigma^0}+G_\Lambda+2\sqrt{3}G_{\ls}\right)=1, \no
%\eea
%which would imply $G_\Lambda( 4 M_\Lambda ^2)\simeq1$, taking into account Eq.~(\ref{eq:uspin-ff}) too.
\\

The aforementioned experimental evidence, i.e. $\eepp$ and $\eell$ are dominated by the Coulomb enhancement
factor and remain almost constant even well above their threshold, has to be tested in the case of 
\be
\ees+s+ . \no
\en 

According to U-spin expectation it should be
\bea
\ees+s+  \sim \sigma(\eepp) \cdot (M_{\pp}/M_{\Sigma^0})^2 \sim 0.53\, \nb. \no
\eea 

This measurement has not yet been done, 
but it is within the \bbr\ or Belle capabilities by means of ISR.

%
% s0s0 and nn form factors
%
\bfi[h!]\vspace{-0mm}
\bc
\epsfig{file=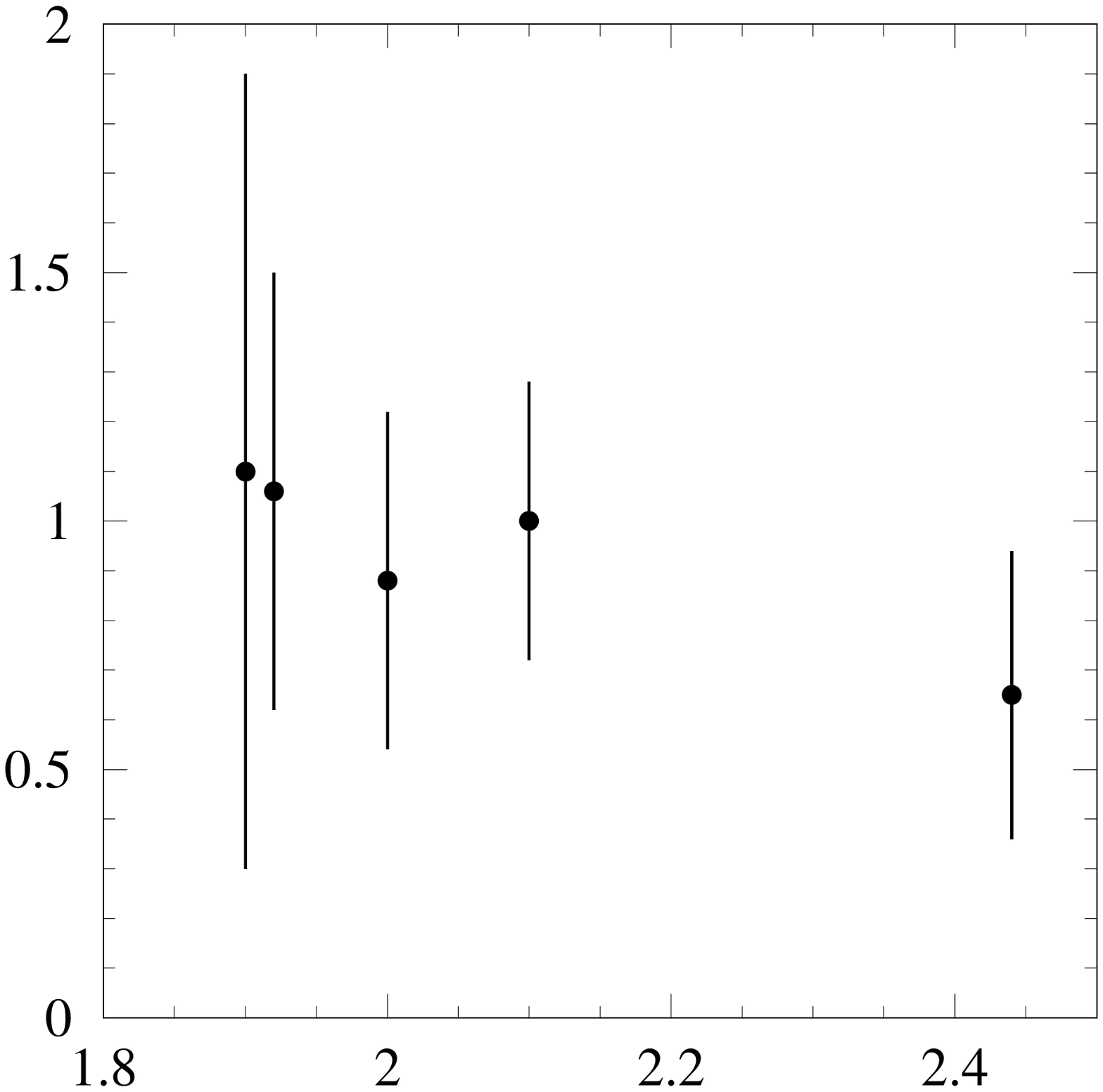,width=80mm}
\put(-60,5){$W_{\nn}$ (GeV)}
\put(-230,135){\rotatebox{90}{$\sigma(\eenn)(\nb)$}}
\vspace{-3mm}
%\bm{80mm}
\caption{\label{fig:s0s0nn}%
The \eenn\ total cross section as measured by
the FENICE Collaboration~\cite{adone}.}%\em
\ec
\efi\vspace{-5mm}
Another important process to understand the nucleon structure is
\be
\eenn\,. \no
\en 
The cross section $\sigma(\eenn)$ has been measured only once, long time ago by the 
FENICE experiment at the $e^+ e^-$ storage ring ADONE~\cite{adone}, that found  
above threshold $\sigma(\eenn)\!\sim$~1~nb, as shown in Fig.~\ref{fig:s0s0nn}.
According to the above mentioned minimal assumption on U-spin invariance it should be
\be
G_n=\frac{3}{2}G_\Lambda-\frac{1}{2}G_{\Sigma^0}, \no
\en
hence
\bea
\sigma(\eenn)=\frac{1}{4}\left(3\sqrt{\sigma_{\ll}}M_\Lambda-
\sqrt{\sigma_{\ss}}M_{\Sigma}\right)^2\frac{1}{M_n^2}=
0.5\pm0.2\,\nb
\label{eq:nn}
\eea
lower than the FENICE results, but not in contradiction because of their large errors,
while the naive expectation
\bea
\sigma(\eenn)=\sigma(\eepp)\left(\frac{Q_d}{Q_u}\right)^2\simeq 0.2\,\nb \no
\eea
is definitely in  disagreement with them.

Unfortunately it is very unlike that \bbr\ or Belle will ever be able to measure this
process by means of ISR. However BESIII at the $\tau$/charm Factory in China and in part
VEPP2000 in Russia can do
that in the c.m. as well as by means of ISR at lower energies.

As mentioned before full 
U-spin symmetry in electromagnetic interactions of members of a SU(3) flavor multiplet should hold
at enough high energy, at least when strange and non-strange mass differences become negligible.
In this limit it is predicted $G^\Lambda \sim - G^{\Sigma^0}$ and $G^\Lambda \sim 0.5\, G^n$.

In Fig.~\ref{fig:llnn} data on magnetic FF's, scaled by the fourth power of 
$\tau_\b=W_{\bb}/2M_\mathcal{B}$ are shown as a function of $\tau_\b$. 
Strange baryon FF's are obtained under the  hypothesis $|G_E^\b|=|G_M^\b|$, 
that of the neutron assuming $|G_E^n|= 0$, while the proton magnetic FF, 
more properly, is achieved by means of dispersion relations using also
the proton angular distribution measurements. The data show a trend 
in agreement with the full U-spin symmetry predictions. 
By the way $\Lambda$ data and U-spin symmetry confirm
the unexpected high cross section $\sigma(\eenn)$, with respect to $\sigma(\eepp)$.
However, data on both $G^{\Sigma^0}$ and $G^n$ are quite poor 
and much better measurements are demanded, in particular in the case of \eenn. 

Various theoretical models and phenomenological descriptions make
predictions on baryon time-like FF~\cite{models}. In particular the \bbr\ cross section, 
angular distributions and \eenn\ cross section have been reproduced, modeling final state
interactions by means of a suitable potential~\cite{milstein}.

  %
% s0s0 and nn form factors
%
\bfi[h]
\bc
\epsfig{file=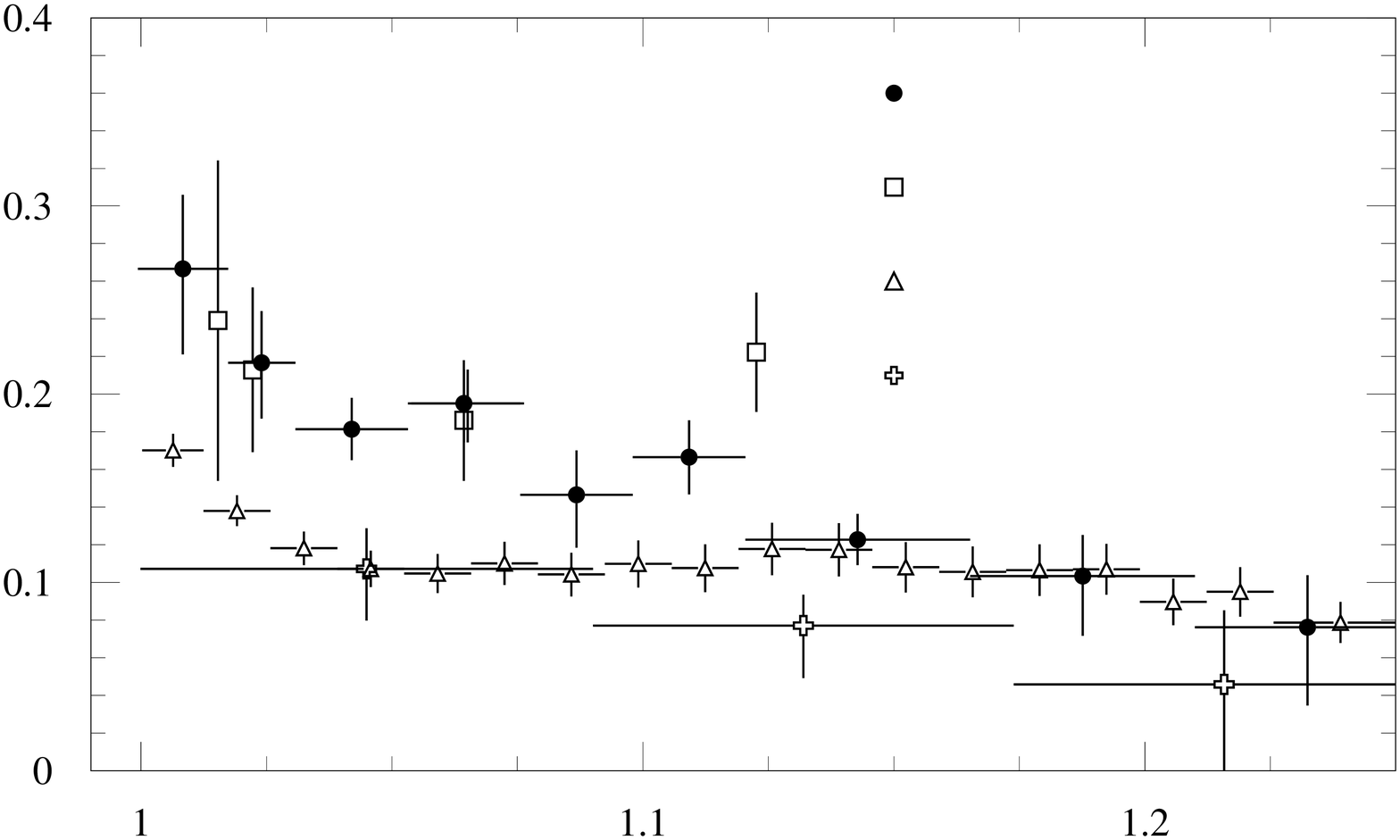,height=80mm}
\put(-85,-10){$\tau_\b=W_{\bb}/2M_\mathcal{B}$}
\put(-137,194){\bbr\ $|G^\Lambda|\times\tau_\mathcal{B}^4$}
\put(-137,169){FENICE $|G^n|\times\tau_\mathcal{B}^4/2$}
\put(-137,144){\bbr\ $|G^p|\times\tau_\mathcal{B}^4$}
\put(-137,119){\bbr\ $|G^{\Sigma^0}|\times\tau_\mathcal{B}^4$}
\put(-395,50){\rotatebox{90}{[Magnetic baryon form factors]$\times\tau_\mathcal{B}^4$}}
\ec
\vspace{-5mm}
\caption{\label{fig:llnn}%
Comparison among  $|G^\Lambda|$, $|G^{\Sigma^0}|$, $|G^p|$ and 
$|G^n|/2$ scaled by the fourth power of the c.m. energy normalized to the 
mass of the final states: $\tau_\b=W_{\bb}/2M_\mathcal{B}$
($\mathcal{B}=\Lambda,\,\Sigma^0,\,n,\,p$). 
}
\efi
\section{Conclusions}
\label{sec:conclu}

All the \eebb\ cross sections, as measured by the \bbr\ Collaboration, do not vanish at threshold.
In the case of \eepp\ this behavior is explained by the \pp\ Coulomb enhancement factor 
and the form factor normalization: $|G^p(4 M^2_p)| \sim 1$, which could be a general feature 
for baryons. This cross section is remarkably flat near threshold. It turns out that S- and 
D-wave have opposite trends, producing this peculiar behavior.
In the case of \eell, as well as \eepp\ the non-vanishing cross section at threshold is consistent with a
valence quark Coulomb enhancement factor.
The \eess\ cross section is quite smaller than the expectation mentioned above and not in
agreement with full U-spin invariance. However a consistent framework
concerning strange baryon FF's is obtained
just requiring the suppression of electromagnetic transitions between U-spin singlet and
triplet.
Neutron and $\Sigma^+$ FF's are demanded to check this new picture of baryon FF's.

%\section*{Acknowledgments}
%
\acknowledgments
We owe special thanks to Stan Brodsky and Yogi Srivastava, who suggested to look more carefully to
Coulomb-like enhancement factors, and warmly acknowledge \bbr\ physicists from the Budker Institute, 
who achieved the aforementioned cross sections. 
%%%%%%%%%%%%%%%%%%%%%%%%%%%%%%%%%%%%%%%%%%%%%%%%%%%%%%%%%%%%%%%%%%%%%%%%%%%%%%%%%%%%%%%%                     
%
%
%
%%%%%%%%%%%%%%%%%%%%%%%%%%%%%%%%%%%%%%%%%%%%%%%%%%%%%%%%%%%%%%%%%%%%%%%%%%%%%%%%%%%%%%%%


\begin{thebibliography}{99}
%0
\bibitem{zichichipp}M.~Conversi, T.~Massam, A.~Zichichi and T.~Muller,
  %``The leptonic annihilation modes of the proton antiproton system at
  %6.8-(GeV/c)**2 timelike four-momentum transfer,''
  Nuovo Cim.\  {\bf 40}, 690 (1965).
  %%CITATION = NUCIA,40,690;%%
%6
\bibitem{zichichi}T.~Massam and A.~Zichichi,
  %``Nucleon electromagnetic form factors - evidence for q-to-the-minus-6th
  %dependence,''
  Lett.\ Nuovo Cim.\  {\bf 1S1}, 387 (1969)
  [Lett.\ Nuovo Cim.\  {\bf 1}, 387 (1969)];\\
  %%CITATION = NCLTA,1,387;%%
F.~Iachello, A.~D.~Jackson and A.~Lande,
  %``Semiphenomenological fits to nucleon electromagnetic form-factors,''
  Phys.\ Lett.\  B {\bf 43}, 191 (1973).
  %%CITATION = PHLTA,B43,191;%%
%
\bibitem{rinaldo}
  R.~Baldini Ferroli  [BaBar Collaboration],
  %``Proton form factors and related processes in BaBar by ISR,''
  Int.\ J.\ Mod.\ Phys.\  A {\bf 21}, 5565 (2006).
  %%CITATION = IMPAE,A21,5565;%%
%4
\bibitem{jlab-mitbates} M.K. Jones {\it et al.}, Phys. Rev. Lett. {\bf 84}, 1398 (2000);\\
	 O. Gayou {\it et al.}, Phys. Rev. C {\bf 64}, 038202 (2001);\\
	 O. Gayou {\it et al.}, Phys. Rev. Lett. {\bf 88}, 092301 (2001);\\
	 B.D. Milbrath {\it et al.}, Phys. Rev. Lett. {\bf 80}, 452 (1998);
	 Phys. Rev. Lett. {\bf 82}, 2221(E) (1999).
%1
\bibitem{pp} B.~Aubert {\it et al.}  [\bbr Collaboration],
  %``A study of e+ e- --> p anti-p using initial state radiation with BABAR,''
  Phys.\ Rev.\  D {\bf 73}, 012005 (2006)
  [arXiv:hep-ex/0512023].
%2
\bibitem{ll} B.~Aubert {\it et al.}  [BABAR Collaboration],
  %``Study of e+e- --> Lambda anti-Lambda, Lambda anti-Sigma0, Sigma0
  %anti-Sigma0 using Initial State Radiation with BABAR,''
  Phys.\ Rev.\  D {\bf 76}, 092006 (2007)
  [arXiv:0709.1988 [hep-ex]].
%3
\bibitem{noi} S.~Pacetti, R.~Baldini, PANDA Workshop, Orsay 2007.
%5
\bibitem{coulomb}
  A.~D.~Sakharov,
  %``INTERACTION OF AN ELECTRON AND POSITRON IN PAIR PRODUCTION,''
  Zh.\ Eksp.\ Teor.\ Fiz.\  {\bf 18}, 631 (1948)
  [Sov.\ Phys.\ Usp.\  {\bf 34}, 375 (1991)].
%7
\bibitem{somm} A.~Sommerfeld, {\it Atombau und Spektralliniem} (Vieweg, Braunschweig, 1944), Vol. 2, p.130.
               J.~Schwinger, {\it Particles, Sources, and Fields}, Vol. III, p. 80. 
%8
\bibitem{yogi}
G.~Castellani, S.~Reucroft, Y.~N.~Srivastava, J.~Swain and A.~Widom,
  %``Final state interactions and the effects of potentials on particle
  %reactions,''
  arXiv:hep-ph/0509089.
  %%CITATION = HEP-PH/0509089;%%
and references therein.
%9
\bibitem{brodsky-cou}
  S.~J.~Brodsky, A.~H.~Hoang, J.~H.~Kuhn and T.~Teubner,
  %``Angular distributions of massive quarks and leptons close to threshold,''
  Phys.\ Lett.\  B {\bf 359}, 355 (1995)
  [arXiv:hep-ph/9508274] and references therein.
%10
\bibitem{Voloshin}
  M.~B.~Voloshin, Mod.\ Phys.\ Lett.\ A {\bf 18}, 1783 (2003);\\
S.~Dubynskiy, A.~Le Yaouanc, L.~Oliver, J.~C.~Raynal and M.~B.~Voloshin,
  %``Isospin breaking in the yield of heavy meson pairs in e+e- annihilation
  %near threshold,''
  Phys.\ Rev.\  D {\bf 75}, 113001 (2007)
  [arXiv:0704.0293 [hep-ph]];\\
G.~P.~Lepage,
  %``Coulomb corrections for UPSILON (4S) $\to$ B anti-B,''
  Phys.\ Rev.\  D {\bf 42}, 3251 (1990);\\
  %%CITATION = PHRVA,D42,3251;%%
 D.~Atwood and W.~J.~Marciano,
  %``RADIATIVE CORRECTIONS AND SEMILEPTONIC B DECAYS,''
  Phys.\ Rev.\  D {\bf 41}, 1736 (1990).
  %%CITATION = PHRVA,D41,1736;%%
%11
\bibitem{coleman} S.~R.~Coleman and S.~L.~Glashow,
  %``Electrodynamic properties of baryons in the unitary symmetry scheme,''
  Phys.\ Rev.\ Lett.\  {\bf 6}, 423 (1961).
%12
\bibitem{park} D.~Park, {\it Introduction to strong interactions} (W. A. Benjamin, Inc., New York, 1966).
%13
\bibitem{adone}
  A.~Antonelli {\it et al.},
  %``The first measurement of the neutron electromagnetic form factors in  the
  %timelike region,''
  Nucl.\ Phys.\  B {\bf 517}, 3 (1998).
  %%CITATION = NUPHA,B517,3;%
%14
\bibitem{models}
M.~A.~Belushkin, H.~W.~Hammer and U.~G.~Meissner,
  %``Dispersion analysis of the nucleon form factors including meson
  %continua,''
  Phys.\ Rev.\  C {\bf 75}, 035202 (2007)
  [arXiv:hep-ph/0608337];
  %%CITATION = PHRVA,C75,035202;%%

C.~Q.~Geng and Y.~K.~Hsiao,
  %``Determination of Nucleon Form factors from Baryonic B Decays,''
  Phys.\ Rev.\  D {\bf 75}, 094005 (2007)
  [arXiv:hep-ph/0606036];
  %%CITATION = PHRVA,D75,094005;%%

J.~Haidenbauer, H.~W.~Hammer, U.~G.~Meissner and A.~Sibirtsev,
  %``On the strong energy dependence of the e+ e- <--> p anti-p amplitude  near
  %threshold,''
  Phys.\ Lett.\  B {\bf 643}, 29 (2006)
  [arXiv:hep-ph/0606064];
  %%CITATION = PHLTA,B643,29;%%

  H.~W.~Hammer,
  %``Nucleon form factors in dispersion theory,''
  Eur.\ Phys.\ J.\  A {\bf 28}, 49 (2006)
  [arXiv:hep-ph/0602121];
  %%CITATION = EPHJA,A28,49;%%

 R.~Bijker and F.~Iachello,
  %``Re-analysis of the nucleon space- and time-like electromagnetic form
  %factors in a two-component model,''
  Phys.\ Rev.\  C {\bf 69}, 068201 (2004)
  [arXiv:nucl-th/0405028];
  %%CITATION = PHRVA,C69,068201;%%

  F.~Iachello and Q.~Wan,
  %``Structure of the nucleon from electromagnetic timelike form factors,''
  Phys.\ Rev.\  C {\bf 69}, 055204 (2004);
  %%CITATION = PHRVA,C69,055204;%%

 J.~P.~B.~de Melo, T.~Frederico, E.~Pace and G.~Salme,
  %``Electromagnetic form factors of hadrons in the space- and time-like
  %regions within a constituent quark model on the light front,''
{\it In the Proceedings of Workshop on $e^+ e^-$ in the 1-GeV to 2-GeV Range: 
Physics and Accelerator Prospects - ICFA Mini-workshp - Working Group
on High Luminosity $e^+ e^-$ Colliders, Alghero, Sardinia, Italy, 10-13 Sep 2003, pp FRWP006}
  [arXiv:hep-ph/0312255];
  %%CITATION = ECONF,C0309101,FRWP006;%%

  F.~Iachello,
  %``Structure of the nucleon from electromagnetic space-like and time-like
  %form factors,''
{\it In the Proceedings of Workshop on $e^+ e^-$ in the 1-GeV to 2-GeV Range: 
Physics and Accelerator Prospects - ICFA Mini-workshp - Working Group
on High Luminosity $e^+ e^-$ Colliders, Alghero, Sardinia, Italy, 10-13 Sep 2003, pp FRWP003}
  [arXiv:nucl-th/0312074];
  %%CITATION = ECONF,C0309101,FRWP003;%%  

A.~Datta and P.~J.~O'Donnell,
  %``A new state of baryonium,''
  Phys.\ Lett.\  B {\bf 567}, 273 (2003)
  [arXiv:hep-ph/0306097];
  %%CITATION = PHLTA,B567,273;

M.~Karliner and S.~Nussinov,
  %``QCD and e+ e- --> baryon + anti-baryon,''
  Phys.\ Lett.\  B {\bf 538}, 321 (2002)
  [arXiv:hep-ph/0202234];
  %%CITATION = PHLTA,B538,321;%%

M.~Karliner,
  %``Towards the resolution of the e+ e- --> anti-N N puzzle,''
  Nucl.\ Phys.\ Proc.\ Suppl.\  {\bf 108}, 84 (2002)
  [arXiv:hep-ph/0112047];
  %%CITATION = NUPHZ,108,84;%%

M.~Karliner,
  %``A possible resolution of the e+ e- -> anti-N N puzzle,''
in {\it Proc. of the $e^+ e^-$ Physics at Intermediate Energies Conference } ed. Diego Bettoni,
{\it In the Proceedings of $e^+ e^-$ Physics at Intermediate Energies, SLAC, 
Stanford, California, 30 Apr - 2 May 2001, pp W10}
  [arXiv:hep-ph/0108106];
  %%CITATION = ECONF,C010430,W10;%%

  J.~R.~Ellis and M.~Karliner,
  %``On electron positron annihilation into nucleon antinucleon pairs,''
  New J.\ Phys.\  {\bf 4}, 18 (2002)
  [arXiv:hep-ph/0108259];
  %%CITATION = NJOPF,4,18;%%

 S.~Dubnicka, A.~Z.~Dubnickova and P.~Weisenpacher,
  %``Non-standard model of the nucleon electromagnetic structure and its
  %predictability,''
  arXiv:hep-ph/0001240;
  %%CITATION = HEP-PH/0001240;%%

R.~Baldini, S.~Dubnicka, P.~Gauzzi, S.~Pacetti, E.~Pasqualucci and Y.~Srivastava,
  %``Nucleon time-like form factors below the N anti-N threshold,''
  Eur.\ Phys.\ J.\  C {\bf 11}, 709 (1999).
  %%CITATION = EPHJA,C11,709;%%
%15
\bibitem{milstein}V.~F.~Dmitriev and A.~I.~Milstein,
  %``Final state interaction effects in N anti-N production near threshold,''
  Nucl.\ Phys.\ Proc.\ Suppl.\  {\bf 162}, 53 (2006)
  [arXiv:nucl-th/0607003].
%
\end{thebibliography}
\end{document}